\documentclass[conference]{IEEEtran}
\makeatletter
\def\ps@headings{%
\def\@oddhead{\mbox{}\scriptsize\rightmark \hfil \thepage}%
\def\@evenhead{\scriptsize\thepage \hfil \leftmark\mbox{}}%
\def\@oddfoot{}%
\def\@evenfoot{}}
\makeatother
\usepackage[left=.64in,right=.64in,top=1in,bottom=.79in]{geometry}

\usepackage{subcaption}
\usepackage{amsmath}
\usepackage{graphicx}

\newcommand{\argmax}{\text{argmax}}
\usepackage[draft]{hyperref}

\title{OS Fingerprinting: New Techniques and a Study of Information Gain and Obfuscation}
\author{Blake Anderson and David McGrew \\ Cisco Systems, Inc. \\ \texttt{\{blake.anderson,mcgrew\}@cisco.com} }
\date{\today}

\begin{document}
\maketitle

\begin{abstract}
Passive operating system fingerprinting reveals valuable information to the defenders of heterogeneous private networks; at the same time, attackers can use fingerprinting to reconnoiter networks, so defenders need obfuscation techniques to foil them.  We present an effective approach for passive fingerprinting that uses data features from TLS as well as the TCP/IP and HTTP protocols in a multi-session model, which is applicable whenever several sessions can be observed within a time window.  In experiments on a real-world private network, our approach identified operating system major and minor versions with accuracies of 99.4\% and 97.5\%, respectively, and provided significant information gain. We also show that obfuscation strategies can often be defeated due to the difficulty of manipulating data features from all protocols, especially TLS, by studying how obfuscation affects our fingerprinting system.  Because devices running unpatched operating systems on private networks create significant vulnerabilities, their detection is critical; our approach achieved over 98\% accuracy at this important goal. 
%We conclude that passive operating system fingerprinting generates useful intelligence for network defenders, especially in the multi-session model, and it is best done with features from application protocols such as HTTP and TLS.
\end{abstract}

\section{Introduction}
%[2, 3, 5, 27]
%[2] Nmap free security scanner, 2004. URL: http://www.insecure.org/nmap/
%[3] Project details for p0f, 2004. URL: http://freshmeat.net/projects/p0f/
%[5] Xprobe official home, 2004. URL: http://www.sys-security.com/html/projects/X.html
%[27] F. Veysset, O. Courtay, and O. Heen. New tool and technique for
% remote operating system fingerprinting, 2002. URL:
% http://www.intranode.com/fr/doc/ring-short-paper.pdf
% original p0f citation: http://seclists.org/bugtraq/2000/Jun/141

Private internal networks face many threats, including attacks from external devices~\cite{Cheswick:2003:FIS:862335}, infected internal devices~\cite{Virvilis:2013:BFW:2545118.2545207}, and unauthorized devices~\cite{minipwner,Wei:2007:POR:1298306.1298357}. One important defensive tool is passive operating system fingerprinting, which identifies the OS of a host solely through the observation of network traffic. This class of techniques originated nearly two decades ago as a way to understand remote devices sending network attack traffic~\cite{honeynet-know}, and was rapidly embraced by the open source community~\cite{p0f-2000}. Passive OS fingerprinting was then further developed by the research community.  Lippmann et.~al.~\cite{Lippmann-2003} introduced the idea of near-match fingerprints, used machine learning classifiers to generate them, and determined the OS categories that are identifiable via fingerprinting. Tyagi et.~al.~\cite{DBLP:journals/ieeesp/TyagiPMT15} used passive OS fingerprinting of TCP/IP to detect unauthorized operating systems on private internal networks. Because vulnerable OS versions are typically present on private networks~\cite{DBLP:conf/ccs/BilgeD12}, another important use of OS fingerprinting is the detection of outdated versions containing vulnerabilities.  

The data features originally used in fingerprinting were from TCP/IP headers, but more recent work has made use of features from HTTP headers~\cite{MBYS11,p0fv3} and unencrypted fields from the TLS/SSL handshake~\cite{httpsInterception2017,DBLP:conf/IEEEares/HusakCJC15}. These features can be analyzed independently when only a single session's data is available, which is not uncommon in some scenarios. In other scenarios, such as passively monitoring an internal network, multiple sessions can be observed, and it pays to utilize features accumulated over time from multiple protocols.  %We describe and study the multi-session model in detail below.  

Although it is beneficial for network administrators to use passive fingerprinting to identify operating systems on their network, attackers have also embraced these techniques to search for potential victims. Because of concerns around this malicious use of identification, defenders have sought ways to use obfuscation to defeat the technique, e.g., by rewriting the
fields in network headers~\cite{conf/cns/AlbaneseBJ15}. These techniques can obfuscate individual sessions or raw data features that a user controls, but they are less successful in the multi-session model, as it is uncommon for a user to have the ability to rewrite all possible network protocols being sent from their device. For example, rewriting TLS metadata can result in failed sessions if the client does not support the rewritten cipher suites, or, as is common in an enterprise setting, a user may not have the appropriate device privileges to modify the installed applications or libraries. This lack of control is a key assumption we make when evaluating our multi-session model in the presence of obfuscation.

A related technique is active OS fingerprinting, in which one or more packets are sent to a device in order to trigger an observable response.  Shu and Lee~\cite{Shu06networkprotocol} formalized passive and active fingerprinting and developed the Parameterized Extended Finite State Machine (PEFSM) to model behavior when multiple messages are sent and received.  Greenwald et.~al.~\cite{Greenwald:2007:TUO:1323276.1323282} studied active fingerprinting, and showed that information gain can be used to minimize the number of probes that are needed. The work presented in this paper is strictly passive; these techniques are less disruptive to networks and applications, are easier to integrate into network monitoring workflows, and they allow for retrospective detection.  %Extending our methods to the active setting would be interesting future work.

Fingerprinting has been generalized to characterize endpoint properties other than operating systems.  Kohno et. al.~\cite{kohno-remote-2005} used passive observations of the TCP Timestamp option to fingerprint individual devices based on their clock skew. To passively fingerprint devices on Industrial Control System networks, Formby et. al. introduced cross-layer response times \cite{DBLP:conf/ndss/FormbySLRB16}. Other works have applied the idea to the identification of particular protocols or applications, e.g., Conti et.~al.~\cite{Conti:2004:PVF:1029208.1029216} fingerprinted attack tools. While we do not specifically address application identification in this paper, the multi-session model that we present is equally applicable because it is common for applications to use multiple protocols.

%To be useful to a network defender, OS fingerprinting must be able to distinguish the OS variants that are actually in use on the network; it offers little value when it cannot distinguish between two equally prevalent operating systems, or when most hosts run the same OS.  In the following, we use both the accuracy of fingerprinting and its information gain to quantify its real-world effectiveness.  

\subsection{Our contribution}
%We provide a formal analysis of the information gain due to fingerprinting using three distinct fingerprinting types: TCP/IP, TLS, and HTTP. 
While TCP/IP and HTTP features have previously been used for passive OS fingerprinting, and TLS features have been used to fingerprint browsers, we describe the first system that integrates all of these data types in a multi-session model to identify the major and minor versions of operating systems. This system accumulates data features within a fixed time window, and applies a machine learning classifier to utilize them. The inclusion of TLS features is crucial because TLS-encrypted HTTPS sessions are increasingly used over unencrypted HTTP sessions. We apply our system to real-world network data collected over a period of 6 days. We compute the information gain due to fingerprinting relative to both single-session and multi-session data features, and show that the latter provides practicable results: on our network, it reduced the uncertainty per host from 2.41 to 0.13 bits.  Significantly, our system can distinguish between minor versions, e.g., \texttt{OSX 10.12.4} versus \texttt{OSX 10.12.3}, with an accuracy of 97.5\%, achieving the fidelity needed to detect vulnerable operating system variants. Finally, we evaluate the multi-session model when a device obfuscates its raw features, and show that the system maintains its accuracy even when the majority of a protocol's features are obfuscated.

\section{Formalization}
In this work, we consider only passive fingerprinting, which we formalize as the process of assigning one or more categories $C$ from a set of categories ${\cal C} = \{ c_1, c_2, \ldots \}$ to some observed network traffic based on a vector of data features $f = \{f_1, f_2, \ldots \}$ by an assignment function $a : {\cal F} \rightarrow {\cal C}^*$, where $\cal F$ denotes the set of possible feature vectors or fingerprints.  Each category can be identified with a descriptive label, such as \texttt{OSX 10.11.6}. A raw feature $r \in \cal R$ is a data element that is extracted directly from a network packet. If the assignment function is non-surjective, i.e., always returns a single category for every possible $f$, it is called \textit{unique}; otherwise, the assignment function is called \textit{non-unique}.

A \textit{fingerprinting scheme}, $({\cal C}, a)$, consists of a set of categories together with an assignment function. Given a fingerprinting scheme, $({\cal C}, a)$, with a non-unique assignment function, it is always possible to find a fingerprinting scheme, $({\cal C}', a')$, with a unique assignment function, $a'$, such that each category in ${\cal C}'$ corresponds to one or more categories in $\cal C$, and the categories associated with $a'(f)$ are exactly those returned by $a(f)$. In these terms, Lippman et. al.~\cite{Lippmann-2003} found a set of categories by merging raw operating system categories that supports a unique assignment function for their features. For instance, when fingerprinting ${\cal C}_\text{OS}$, it may be impossible to distinguish between \texttt{BSD} and \texttt{OSX}, and thus a unique fingerprinting scheme must include the composite category $\{\texttt{BSD},\texttt{OSX}\}$.  In general, $\cal C$ should contain all categories of interest, even though it may not be possible to find a unique assignment function.

\subsection{Prior Probabilities and Maximum Likelihood}
Before the fingerprinting process, the probability that a particular category $c$ will occur is given by the prior distribution $P(C)$, where $C$ is a random variable associated with the unknown category. When the assignment function in a fingerprinting scheme is not unique, and it is necessary or desirable to work with a single category rather than a set of likely categories or a distribution over categories, the optimal strategy is to choose the most likely category $c = \argmax_{c^* \in a(f)} p(c^*)$. This maximum likelihood rule  creates a unique assignment function by selecting the category with the largest probability from a composite category. For instance, in most network environments $p(\texttt{OSX}) > p(\texttt{BSD})$ for mobile computers, and thus \texttt{OSX} would be chosen by the maximum likelihood rule.

\subsection{Information Gain}
Fingerprinting reduces the uncertainty by eliminating some possibilities, and the information gained by a fingerprinting process can be quantified by treating the actual category, $c$, the fingerprint, $f$, and the set of categories returned by $a(f)$, as random variables. $p(f \mid c)$ is the conditional probability that fingerprint, $f$, will be observed given the category, $c$, and $p(c \mid f)$ is the posterior probability. When the assignment function is unique, these probabilities are given by:
\begin{align}
p(f \mid c) & = \begin{cases}
			1 & \text{if } c \in a(f) \\
			0 & \text{otherwise},
		\end{cases} \\
p(c \mid f) & = \begin{cases}
			\frac{P(c)}{ N_f} & \text{if } c \in a(f) \\
			0 		  & \text{otherwise,}
		\end{cases}
\end{align}
where the normalization factor $N_f = \sum_{c^* \in a(f) } p(c^*)$ apportions probability among all of the categories in $a(f)$. The second equation follows directly from Bayes' rule. Before the fingerprinting process, the distribution of categories is given by the prior, $P(C)$, and after observing a particular fingerprint, $f$, the probability is the posterior as above.

The prior distribution captures all of the knowledge about the probability with which categories occur. If there is little uncertainty in that distribution, e.g., $p(\texttt{Windows}) = 0.99$, then there is little information gained on average. Alternatively, if the prior distribution has high uncertainty, e.g., $p(\texttt{Windows}) = p(\texttt{OSX}) = 0.5$, then fingerprinting techniques can offer a significant advantage.  This intuition can be formalized in terms of Shannon entropy. Let $C$ and $F$ be random variables representing the categories and the fingerprints, respectively. Before fingerprinting is applied, the entropy of the category distribution is $H(C)$, and after fingerprinting is applied, the entropy of the category distribution is $H(C \mid F)$, where:
\begin{align}
\label{equation:entropy_before}
H(C)            & = -\sum_{c \in C} p(c) \log p(c) \\
H(C \mid F) & = -\sum_{f \in F} p(f) \sum_{c^* \in a(f)} p(c^* | f) \log p(c^* | f)
\end{align}
The information gain of applying fingerprinting is then: %defined as:
\begin{align}
I(C;F) & = H(C) - H(C \mid F)
\label{equation:information_gain}
\end{align}

Equation \ref{equation:information_gain} can be equivalently stated in terms of KL-Divergence:
\begin{align}
I(C;F) & = \sum_{f \in F} p(f) D_\text{KL} \left(p(c \mid f)  \hspace{.4mm}\Vert\hspace{.4mm}  p(c)\right)
\label{equation:information_gain_kl}
\end{align}
where $D_\text{KL} \left(p(c | f)  \hspace{.4mm}\Vert\hspace{.4mm}  p(c)\right)$ is the per-fingerprint information gain, and is defined as:
\begin{align}
D_\text{KL} \left(p(c | f) \hspace{.1mm}\Vert\hspace{.1mm} p(c)\right) & = \sum_{c^* \in a(f)} p(c^* | f) \log \frac{p(c^* | f)}{p(c^*)}
\label{equation:information_gain_per_feature}
\end{align}

\bgroup
\def\arraystretch{1.2}
\begin{table}
\center
\begin{tabular}{l|c|c|c}
\hline
Data Type & $H(C)$ & $H(C \mid F)$ & $I(C;F)$ \\
\hline
All - Multi & 2.41 & 0.13 & 2.28 \\
TCP/IP - Multi & 2.48 & 1.36 & 1.12 \\
TLS - Multi & 2.47 & 0.31 & 2.16 \\
HTTP - Multi & 2.42 & 0.37 & 2.06 \\
TCP/IP & 2.44 & 1.42 & 1.02 \\
TLS & 2.56 & 0.74 & 1.82 \\
HTTP & 2.49 & 0.35 & 2.14 \\
\hline
\end{tabular}
\caption{The information gain due to fingerprinting. The ``Multi" data types use the multi-session, windowed approach described in Section \ref{section:results:multi_flow}.}
\label{table:information_gain}
\end{table}
\egroup

\section{Raw Data Features}
\label{features}
Our experiments are based on three different types of network data: TCP/IP, TLS, and HTTP. Table \ref{table:information_gain} provides an overview of the information gain for each of the data types, computed using Equation \ref{equation:information_gain}. The TLS and HTTP fingerprints perform the best in our experiments, and this is reflected in Table \ref{table:information_gain}. %The variability of $H(C)$ is due to endpoints' variable use of each protocol. 
Here, $H(C)$ is the empirical entropy estimate based on the distribution of the data feature.
The ``Multi" data types are included for completeness, and are constructed by concatenating all fingerprinting information contained within a 60 minute window and using a machine learning classifier as described in Section \ref{section:results:multi_flow}. 

The protocol-specific data features have the following form: an ordered list of elements, each of which is either a type code or a (type code, string) pair.  The type codes are unsigned integers or distinguishing strings used in the protocol, which we treat as categorical variables.  The list is constructed by populating its elements in the order in which the type codes are observed in the session.  For each protocol, there is a set of type code values  for which a string is obtained from the observed packet and included in the list as a (type code, string) pair; for all other type code values, only that code is included. 

\subsection{TCP/IP}

The earliest passive OS fingerprinting schemes used features from TCP/IP, such as the IP Time To Live (TTL) values, the list of TCP option types, and the values of the Maximum Segment Size (MSS) and Window Scale (WS) options. For our experiments, we collected the TTL and the ordered list of all TCP options associated with a TCP SYN packet. Repeated options were allowed in the ordered list. We also included the data for the MSS and WS TCP options in the ordered list. In total, we found 61 unique TCP/IP fingerprints in our dataset.

\subsection{TLS}

The TLS protocol \cite{rfc5246} has a complex set of protocol options that can be chosen by the client.  The TLS fingerprinting features that we use are taken from the TLS \texttt{ClientHello} message and include the cipher suite offer vector, which is an ordered list of cipher suite type codes, and the extensions vector, which is an ordered list of extension type codes. Similar to the TCP MSS/WS options, we collected the data associated with the following extensions:
\begin{itemize}
\item \texttt{supported\_groups}
\item \texttt{ec\_point\_formats}
\item \texttt{application\_layer\_protocol} \newline
\texttt{\_negotiation}
\end{itemize}
The \texttt{supported\_groups} and \texttt{ec\_point\_formats} extensions \cite{rfc4492,rfc7919} provide the cryptographic parameters that the client supports. The \texttt{application\_layer\_protocol} \texttt{\_negotiation} extension \cite{rfc7301} provides an ordered list of application-layer protocols that the client supports, e.g., \texttt{HTTP/2}, \texttt{HTTP/1.1}, etc. There were 1,054 unique TLS fingerprints in our dataset.

\subsection{HTTP}

\bgroup
\def\arraystretch{1.2}
\begin{table}
\center
\begin{tabular}{l|c}
\hline
HTTP Fingerprint & Information \\
 & Gain \\
\hline
\texttt{Windows NT 6.1; WOW64;} & 0.1494 \\
\texttt{Windows NT 10.0; WOW64;} & 0.1358 \\
\texttt{Intel Mac OS X 10.12;} & 0.1158 \\
\hline
\texttt{Microsoft NCSI} & 0.0023 \\
\texttt{WinHttpClient} & 0.0003 \\
\texttt{iPhone8,1/10.3.1} & 0.0001 \\
\hline
\end{tabular}
\caption{A summary of the information gain due to individual HTTP fingerprints. All three high information gain fingerprints are \texttt{Firefox/52.0}, but edited for clarity.}
\label{table:information_gain_http}
\end{table}
\egroup

\begin{table*}
\center
\begin{tabular}{l|r|r|r|r|r|r||r}
\hline
Operating System   & Day$_0$ & Day$_1$ & Day$_2$ & Day$_3$ & Day$_4$ & Day$_5$ & Total \\
\hline
Win 10.0.1439 & 71 (264,405) & 57 (172,005) & 41 (147,410) & 62 (201,390) & 82 (251,789) & 74 (225,285) & 387 (1,262,284) \\
Win 10.0.1058 & 360 (1,315,011) & 261 (1,483,997) & 160 (1,259,831) & 246 (1,782,281) & 346 (1,583,422) & 371 (1,654,996) & 1,744 (9,079,538) \\
Win 6.3.960 & 1 (979) & 1 (1,068) & 0 (0) & 1 (4,015) & 0 (0) & 2 (2,094) & 5 (8,156) \\
Win 6.1.760 SP1 & 466 (2,350,465) & 349 (2,614,222) & 175 (2,648,420) & 320 (2,489,747) & 436 (2,582,081) & 501 (2,462,193) & 2,247 (15,147,128) \\
OSX 10.12.5 & 1 (2,283) & 3 (1,582) & 1 (846) & 3 (7,908) & 3 (8,706) & 4 (13,637) & 15 (34,962) \\
OSX 10.12.4 & 141 (384,379) & 81 (322,412) & 60 (435,924) & 88 (489,643) & 120 (444,859) & 164 (657,403) & 654 (2,734,620) \\
OSX 10.12.3 & 67 (139,407) & 41 (174,649) & 24 (198,534) & 35 (165,888) & 54 (106,140) & 47 (99,044) & 268 (883,662) \\
OSX 10.12.2 & 9 (8,258) & 3 (15,133) & 8 (25,441) & 3 (3,296) & 3 (4,885) & 11 (9,659) & 37 (66,672) \\
OSX 10.12.1 & 5 (14,343) & 2 (9,767) & 1 (19) & 1 (4,368) & 6 (16,072) & 14 (30,967) & 29 (75,536) \\
OSX 10.12.0 & 3 (38,337) & 4 (51,953) & 1 (16,165) & 2 (21,736) & 5 (12,385) & 6 (11,183) & 21 (151,759) \\
OSX 10.11.6 & 245 (617,126) & 199 (986,109) & 116 (743,465) & 183 (896,880) & 228 (874,516) & 322 (868,068) & 1,293 (4,986,164) \\
OSX 10.11.5 & 6 (20,357) & 4 (47,055) & 1 (719) & 1 (108) & 4 (953) & 4 (3,256) & 20 (72,448) \\
OSX 10.11.4 & 2 (7,691) & 0 (0) & 1 (17,502) & 1 (10,071) & 1 (7,511) & 2 (9,905) & 7 (52,680) \\
OSX 10.11.3 & 0 (0) & 0 (0) & 1 (3,486) & 4 (7,277) & 1 (8,924) & 0 (0) & 6 (19,687) \\
OSX 10.11.1 & 1 (1,705) & 0 (0) & 0 (0) & 0 (0) & 0 (0) & 0 (0) & 1 (1,705) \\
OSX 10.11.0 & 2 (305) & 0 (0) & 0 (0) & 0 (0) & 1 (14,596) & 0 (0) & 3 (14,901) \\
OSX 10.10.5 & 42 (159,741) & 36 (145,468) & 21 (171,763) & 32 (169,626) & 33 (99,781) & 33 (175,077) & 197 (921,456) \\
OSX 10.9.5 & 9 (12,581) & 9 (6,763) & 0 (0) & 5 (5,541) & 5 (4,244) & 1 (2,249) & 29 (31,378) \\
iOS 10.3.1 & 0 (0) & 7 (2,662) & 3 (286) & 13 (2,590) & 13 (4,238) & 27 (4,438) & 63 (14,214) \\
iOS 10.3 & 8 (1,982) & 2 (1,209) & 4 (1,346) & 4 (600) & 3 (1,402) & 2 (595) & 23 (7,134) \\
iOS 10.2.1 & 45 (18,423) & 36 (13,808) & 14 (6,581) & 21 (7,296) & 37 (18,266) & 57 (20,838) & 210 (85,212) \\
iOS 10.2 & 2 (143) & 3 (447) & 0 (0) & 0 (0) & 3 (297) & 3 (699) & 11 (1,586) \\
iOS 10.1.1 & 6 (4,077) & 6 (1,755) & 0 (0) & 0 (0) & 1 (340) & 0 (0) & 13 (6,172) \\
iOS 10.0.2 & 2 (554) & 1 (1,224) & 1 (75) & 0 (0) & 3 (828) & 2 (73) & 9 (2,754) \\
iOS 9.3.5 & 1 (155) & 2 (273) & 0 (0) & 0 (0) & 0 (0) & 2 (199) & 5 (627) \\
android 7.0 & 0 (0) & 1 (3) & 0 (0) & 0 (0) & 0 (0) & 0 (0) & 1 (3) \\
\hline
Total & 1,495 (5,362,707) & 1,108 (6,053,564) & 633 (5,677,813) & 1,025 (6,270,261) & 1,388 (6,046,235) & 1,649 (6,251,858) & 7,298 (35,662,438) \\
\hline
\end{tabular}
\caption{An overview of the diversity of operating systems in the network that we monitored. The number of unique hosts as well as the number of flows is given as: hosts (flows).}
\label{table:environment_overview}
\end{table*}

For the HTTP fingerprint, we only use the \texttt{User-Agent} header value, with capitalization maintained. Most browsers will include some informative string indicating the operating system within the \texttt{User-Agent}, e.g., \texttt{Windows NT 10.0}. As Table \ref{table:information_gain} illustrates, HTTP fingerprints had the most information gain in a single-session setting. Despite their high information gain, \texttt{User-Agent} strings should be used with caution because they are easy to manipulate, and, in most cases, their modification would not cause failures. In our dataset, there was a total of 1,848 unique HTTP fingerprints.

To provide some intuition about what characteristics a fingerprint with high information gain has, Table \ref{table:information_gain_http} lists examples of HTTP fingerprints with high and low normalized information gain. These values were normalized by multiplying the fingerprint's probability, $p(f)$, with the fingerprint's information gain (from Equation~\ref{equation:information_gain_per_feature}). The high information gain features appear frequently within our dataset, and also have a strong association with one or a few operating system types. The low information gain features either appear infrequently or are equally probable across a large number of operating system types. \texttt{Microsoft NCSI} and \texttt{WinHttpClient} both appear frequently in our dataset, but in different Windows versions proportional to those versions overall probabilities. \texttt{iPhone8,1/10.3.1} is only seen with a single operating system type, but also only appears once in our dataset.

Similar to the previous data types, we did experiment with a representation that included the full, ordered list of HTTP headers and the data associated with the \texttt{User-Agent}. This led to poor results, most likely because each HTTP client can potentially have hundreds of unique lists of HTTP headers, and there wasn't enough data to properly condition the models. For example, there were 138 unique HTTP header combinations for a single \texttt{Firefox/52.0}, \texttt{Win 6.1.760 SP1} \texttt{User-Agent} string.

\bgroup
\def\arraystretch{1.2}
\begin{table}[t!]
\center
\begin{tabular}{l|r|r}
\hline
OS Type & Hosts & Flows \\
\hline
Windows & 4,383 & 25,497,106 \\
OS X & 2,580 & 10,047,630 \\
iOS & 334 & 117,699 \\
Android & 1 & 3 \\
\hline
\end{tabular}
\caption{A summary of the observed network activity by major operating system types.}
\label{table:environment_overview_gen}
\end{table}
\egroup

\begin{figure*}
\centering
\begin{subfigure}[b]{15.0em}
   \includegraphics[scale=0.25]{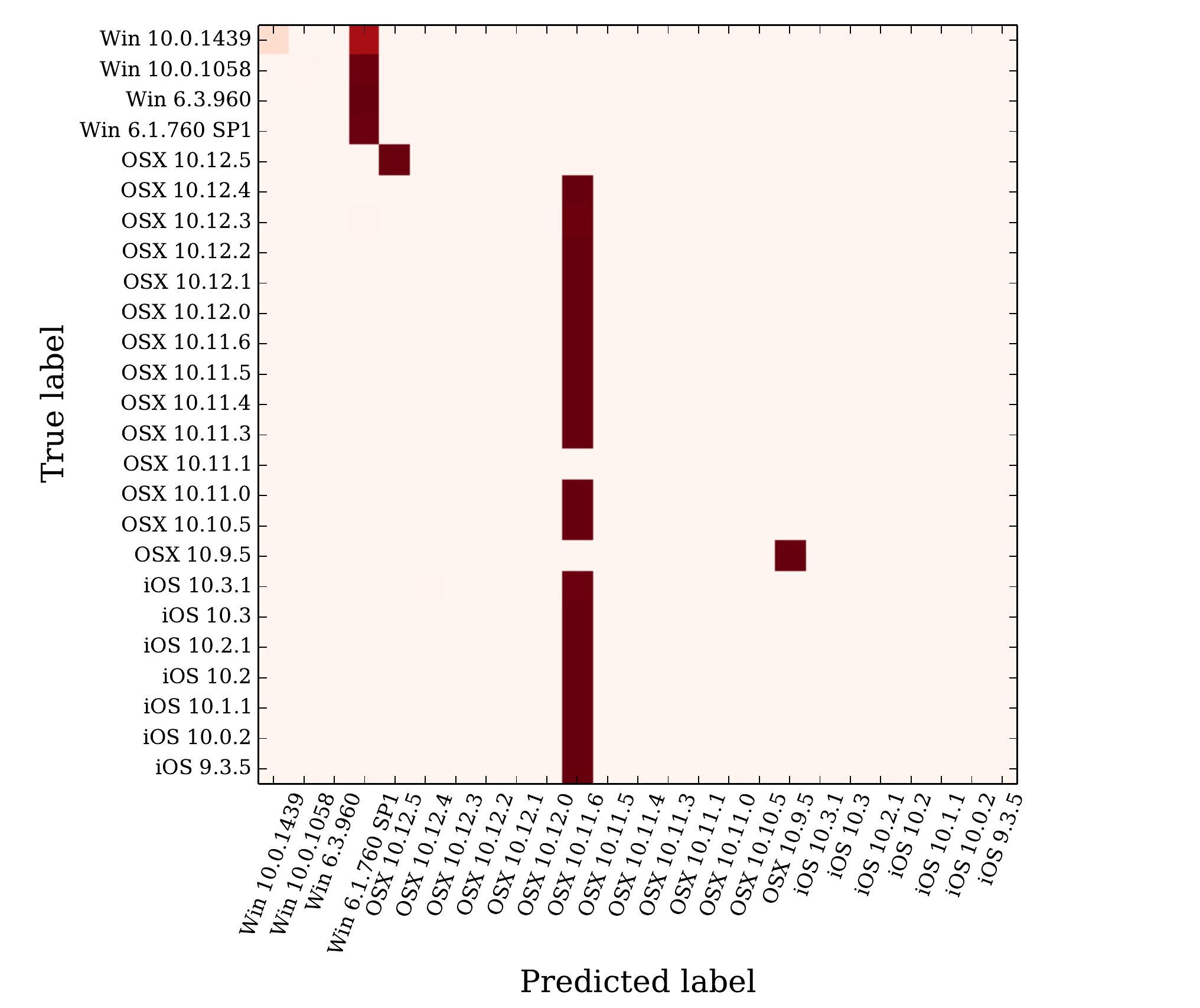}
   \caption{TCP/IP, Accuracy=55.54\%}
   \label{fig:single_flow_conf_tcp}
\end{subfigure}
\begin{subfigure}[b]{15.0em}
   \includegraphics[scale=0.25]{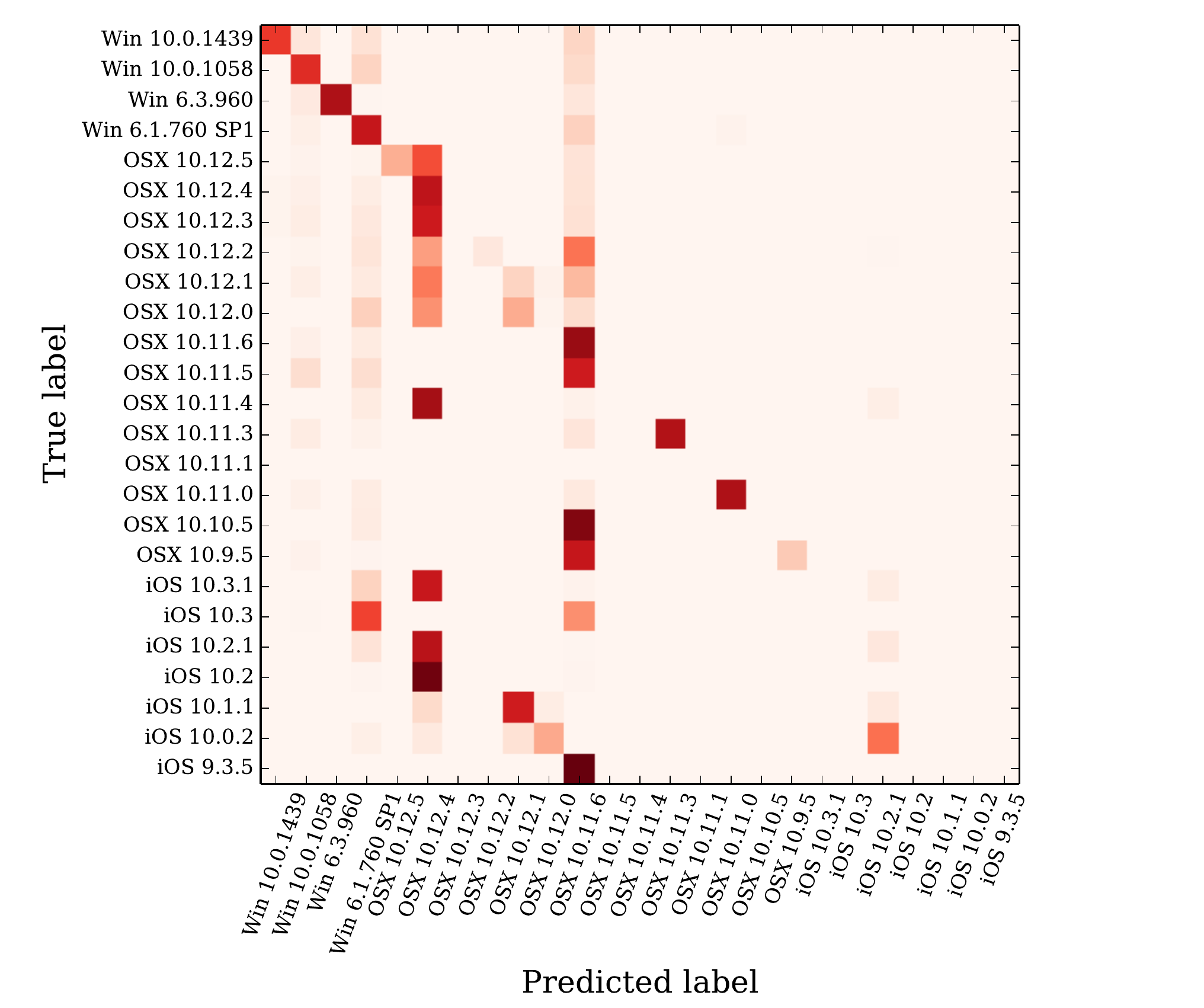}
   \caption{TLS, Accuracy=76.39\%}
   \label{fig:single_flow_conf_tls}
\end{subfigure}
\begin{subfigure}[b]{15.0em}
   \includegraphics[scale=0.25]{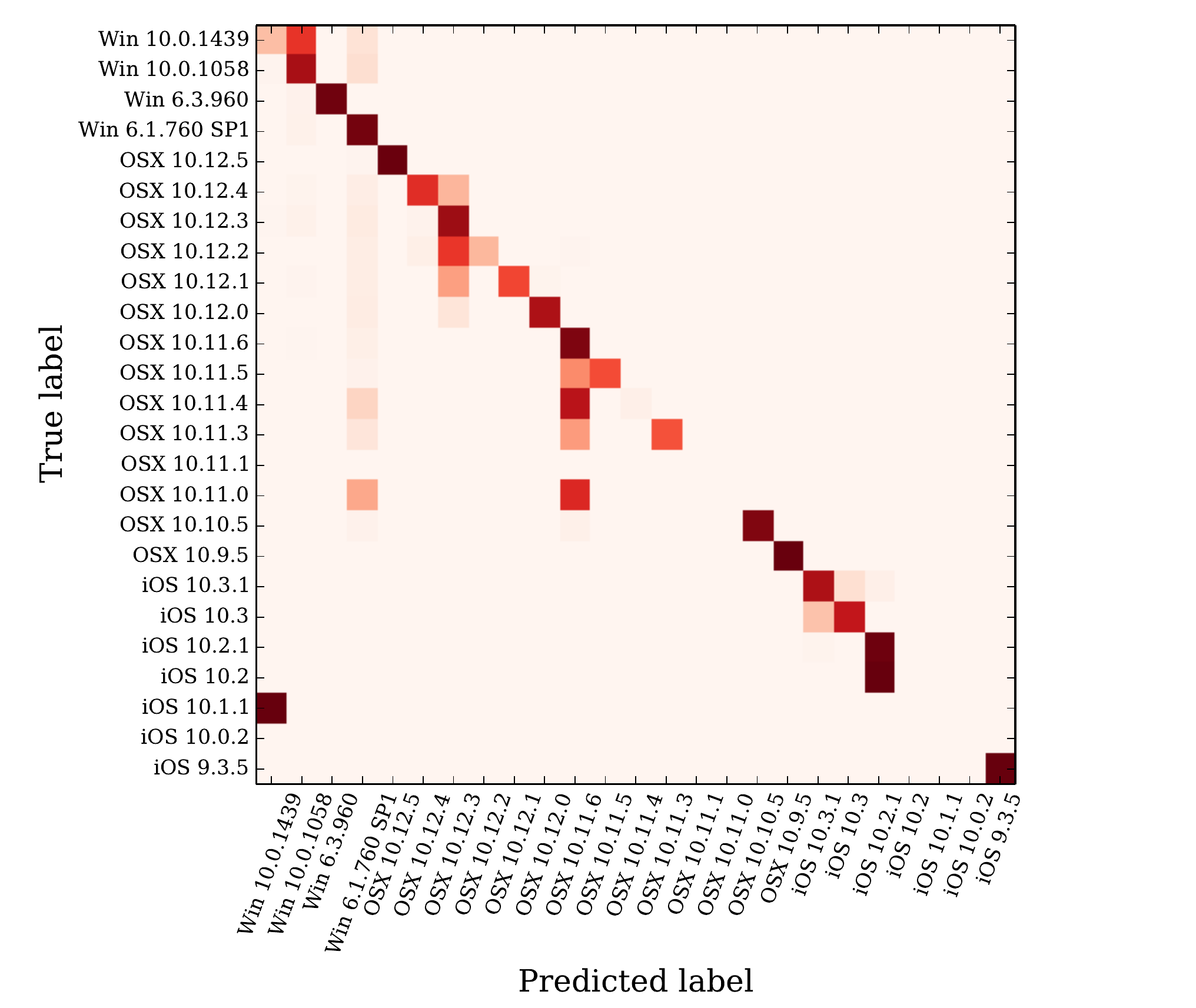}
   \caption{HTTP, Accuracy=87.41\%}
   \label{fig:single_flow_conf_http}
\end{subfigure}
\caption{Confusion matrices for the single session, $\argmax$ classifier.}
\label{fig:single_flow_conf}
\end{figure*}

\section{Collection Environment}
\label{section:environment}
We monitored the Internet connection of an enterprise network with $\sim$3,000 active endpoints for six days in April, 2017.  The internal network was connected to WiFi-attached hosts and the trusted (non-encrypted) interface of a Virtual Private Network (VPN) concentrator.  By accessing the logs of the VPN concentrator, we obtained ground truth information on the OS of the hosts using that device.  The data features described in Section~\ref{features} were obtained for each session using an open source package that we extended for that purpose~\cite{joy}, which monitors a SPAN port and stores a JSON record of each session that includes all of the relevant features from TCP SYN packets, TLS \texttt{ClientHello} messages, and HTTP requests.  The IP addresses and identifying usernames contained within HTTP requests were anonymized, and the anonymized/deanonymized IP address pairs were kept separate from the network flow data as described in Section \ref{section:ethical}.

%We monitored an enterprise network with $\sim$3,000 active endpoints for six days in April, 2017. All of the traffic from the network was collected via a SPAN port and a virtual machine on the network's DMZ. The raw network traffic was then processed using an open source tool: Joy \cite{joy}. Joy was able to collect all relevant features from TCP SYN packets, TLS \texttt{ClientHello} messages, and HTTP requests. Joy stored the data in a convenient JSON format that facilitated further processing. The IP addresses and identifying usernames contained within HTTP requests were anonymized, and the anonymized/deanonymized IP address pairs were kept separate from the network flow data as described in Section \ref{section:ethical}.

Approximately half of the endpoints on the enterprise network were connected using a supported VPN client that logs endpoint information to a central repository. Table \ref{table:environment_overview} provides an overview of the diversity of the operating systems that we observed, as well as the number of hosts and network flows for each operating system type. There were 26 unique operating systems during the six day period, but some had very little data, e.g., \texttt{iOS 9.3.5}. Despite the severe class imbalance, the only OS type that we discarded in all of our experiments was \texttt{android 7.0}. The dominant categories were unsurprising: 1) \texttt{Win 6.1.760 SP1} with 2,247 hosts and 15,147,128 network flows, 2) \texttt{Win 10.0.1058} with 1,744 hosts and 9,079,538 network flows, and 3) \texttt{OSX 10.11.6} with 1,293 hosts and 4,986,164 network flows. Table \ref{table:environment_overview_gen} summarizes Table \ref{table:environment_overview} by combining major operating system types.

Again, the endpoints for which we had ground truth were all using a VPN client to connect to the enterprise network. The VPN tunnel adversely affected the maximum segment size, with the largest MSS observed being 1,260. This undoubtedly limited the efficacy of the TCP/IP fingerprints, and we view the TCP/IP results as a lower bound of the true efficacy. We leave collecting additional datasets and further analyzing TCP/IP fingerprints as future work. We did verify that the VPN client and other relevant network proxies had no effect on either the TLS or HTTP data features.

\section{Results}
\label{section:results}

In all of our experiments, we use data from the first three days of our dataset to train the operating system classification models, and then use data from the last three days to test the models. The operating system label set is taken from Table \ref{table:environment_overview}, but can have some variability in the different experiments due to an insufficient number of relevant network flows appearing in the training data. For example, if there were no HTTP \texttt{User-Agent} strings from \texttt{android 7.0} devices in the training dataset, then we would omit the \texttt{android 7.0} label and any corresponding network traffic from the experiment.

We used a custom program when analyzing network flows in isolation, and used the random forest module from \texttt{scikit-learn} \cite{sklearn} when analyzing features accumulated during time windows. Because of the highly imbalanced nature of our dataset, confusion matrices are presented for the first single and multi-session flow experiments to illustrate the models' ability to learn the minority classes. Each element of the confusion matrix represents the number of samples with label $l_t$ that are classified as label $l_p$. Correctly classified samples, $l_t = l_p$, are the diagonal elements, and misclassified samples, $l_t \neq l_p$, are the off-diagonal elements. As a baseline, a majority-class classifier would have $\sim$35\% accuracy, and a classifier that only correctly labels \texttt{Win 6.1.760 SP1}, \texttt{Win 10.0.1058}, and \texttt{OSX 10.11.6} samples would have $\sim$80\% accuracy.

\subsection{Single Session Model}
\label{section:results:single_flow}

In these experiments, we treated each flow and fingerprint type as independent. The last three days of data were used for testing, and if a network flow in that dataset had a fingerprint, $f$, for the fingerprint type being analyzed, then we selected the operating system with:
\begin{equation}
c = \argmax_{c^* \in a(f)} p(c^*|f)
\end{equation}
where $p(c^*|f)$ was conditioned on the first three days of data in our dataset.

We trained three single session classifiers, one for each fingerprinting type, and Figure \ref{fig:single_flow_conf} presents the confusion matrices for the TCP/IP, HTTP, and TLS fingerprints. Some classes were not in the testing dataset, e.g., \texttt{OSX 10.11.1}, and those rows will be blank in the confusion matrix. TCP/IP performed the worse with a total accuracy of 55.54\%. As Figure \ref{fig:single_flow_conf_tcp} shows, TCP/IP almost exclusively separates the traffic into \texttt{Win 6.1.760 SP1}, the dominating Windows OS, and \texttt{OSX 10.11.6}, the dominating Mac OS.

Both TLS and HTTP had better performance, with HTTP's accuracy at 87.41\%. TLS had some similar behavior to TCP/IP, favoring \texttt{Win 6.1.760 SP1}, \texttt{OSX 10.12.4}, and \texttt{OSX 10.11.6}. HTTP's performance is not surprising because 8 out of the 10 most frequent HTTP \texttt{User-Agent} strings contained at least the major version number of the operating system, e.g., 10.12 for Mac or 10.0 for Windows.

\subsection{Multi-Session Model}
\label{section:results:multi_flow}

\bgroup
\def\arraystretch{1.2}
\begin{table}
\center
\begin{tabular}{l|c}
\hline
Data Type & Number of \\
 & Fingerprints \\
\hline
TCP/IP Fingerprints & 1.72 \\
TLS Fingerprints & 10.25 \\
HTTP Fingerprints & 3.35 \\
All Fingerprints & 13.97 \\
\hline
\end{tabular}
\caption{The average number of unique fingerprints observed in each 60 minute window.}
\label{table:fingerprints_per_window}
\end{table}
\egroup

\begin{figure*}
\centering
\begin{subfigure}[b]{23em}
   \includegraphics[scale=0.36]{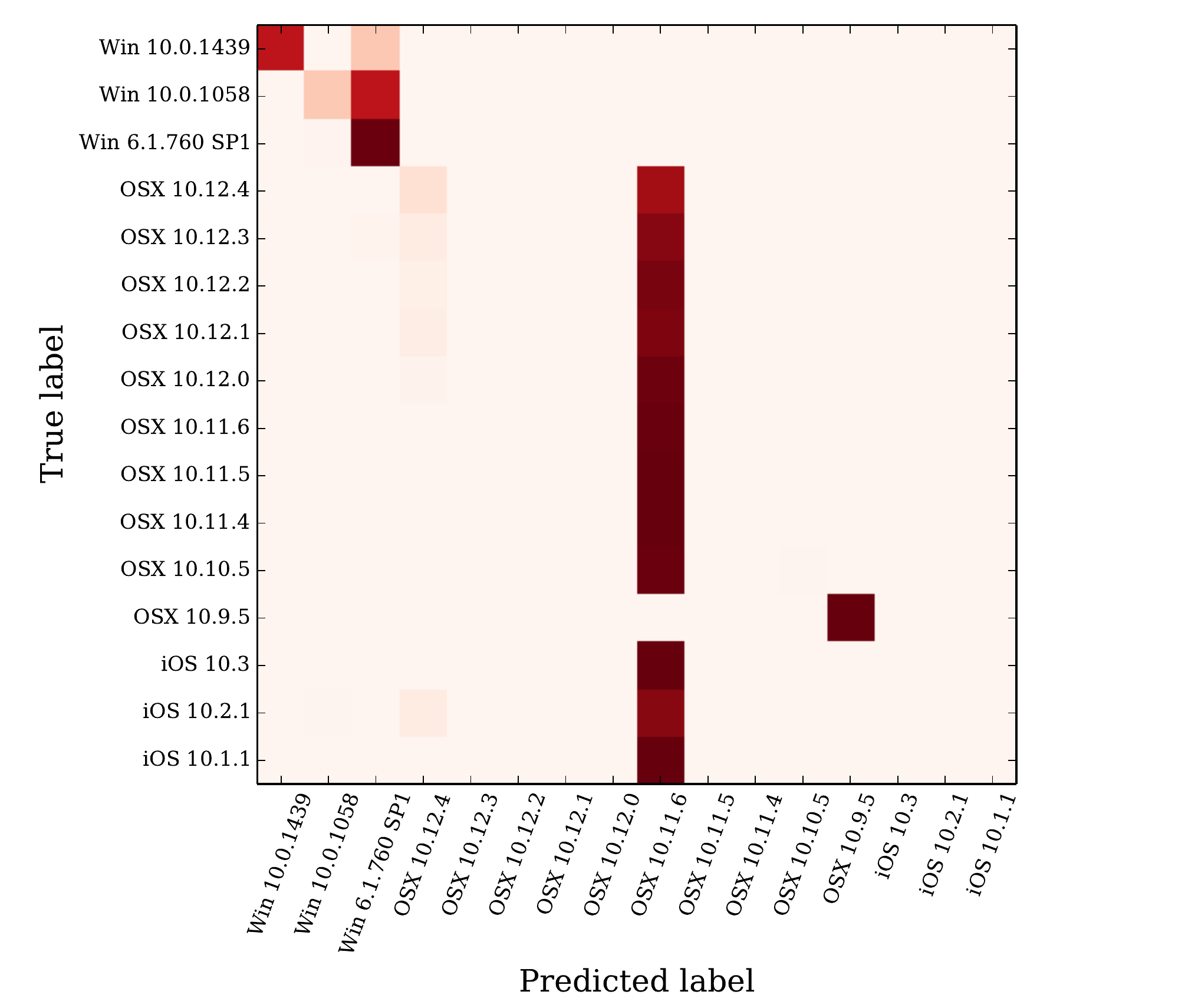}
   \caption{TCP/IP, Accuracy=62.30\%}
   \label{fig:multi_flow_conf_tcp}
\end{subfigure}
\begin{subfigure}[b]{23em}
   \includegraphics[scale=0.36]{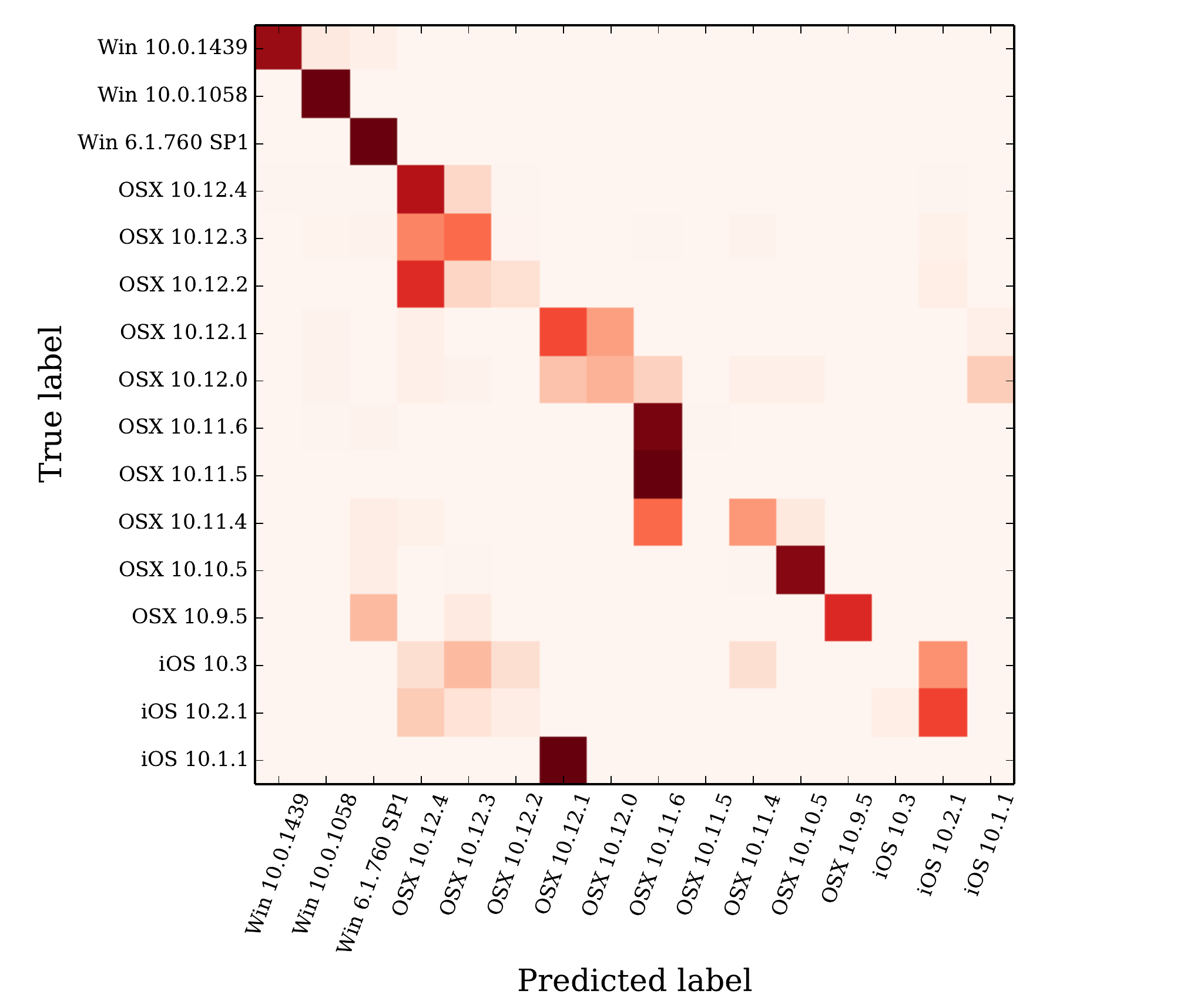}
   \caption{TLS, Accuracy=93.67\%}
   \label{fig:multi_flow_conf_tls}
\end{subfigure}
\begin{subfigure}[b]{23em}
   \includegraphics[scale=0.36]{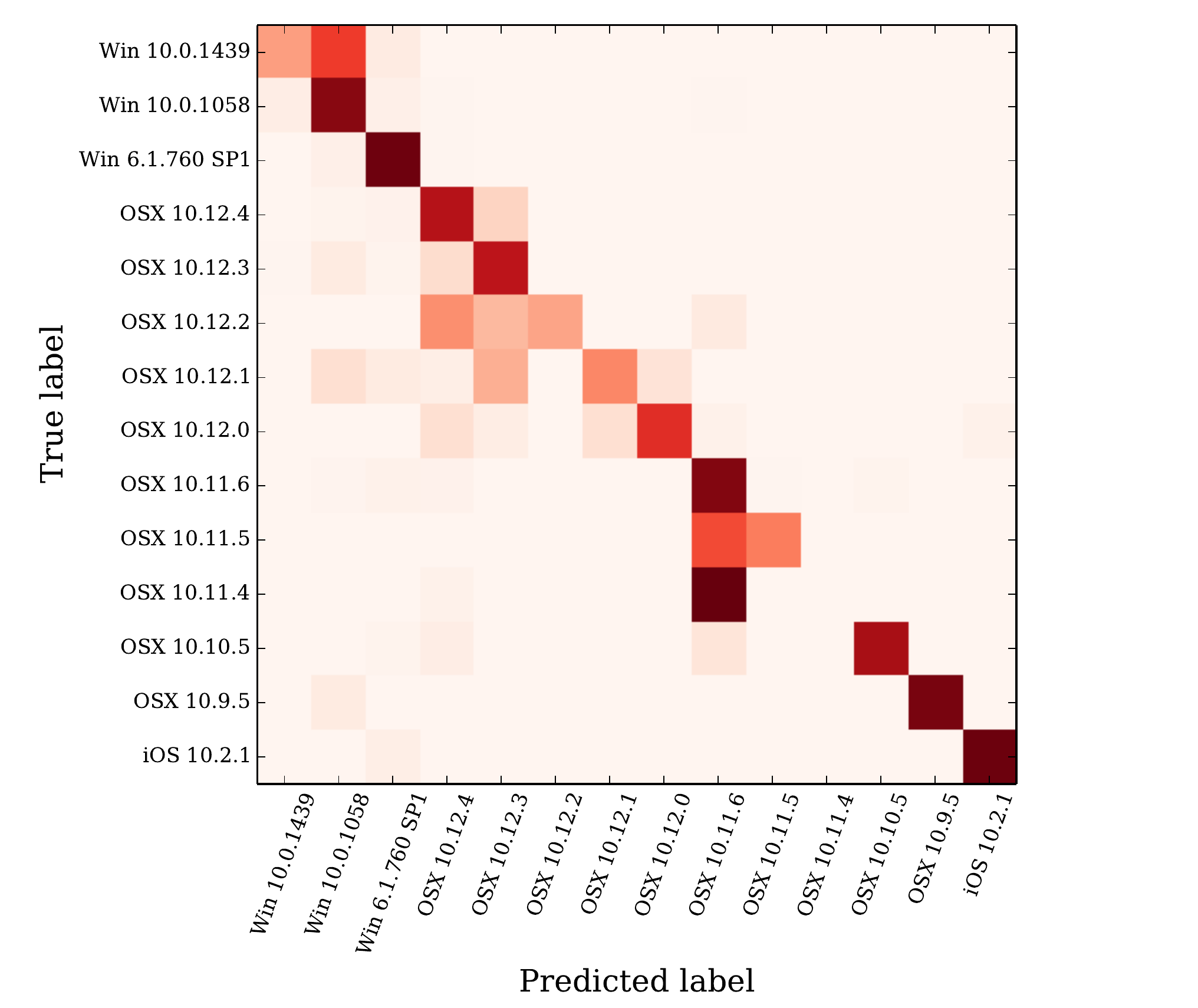}
   \caption{HTTP, Accuracy=87.69\%}
   \label{fig:multi_flow_conf_http}
\end{subfigure}
\begin{subfigure}[b]{23em}
   \includegraphics[scale=0.36]{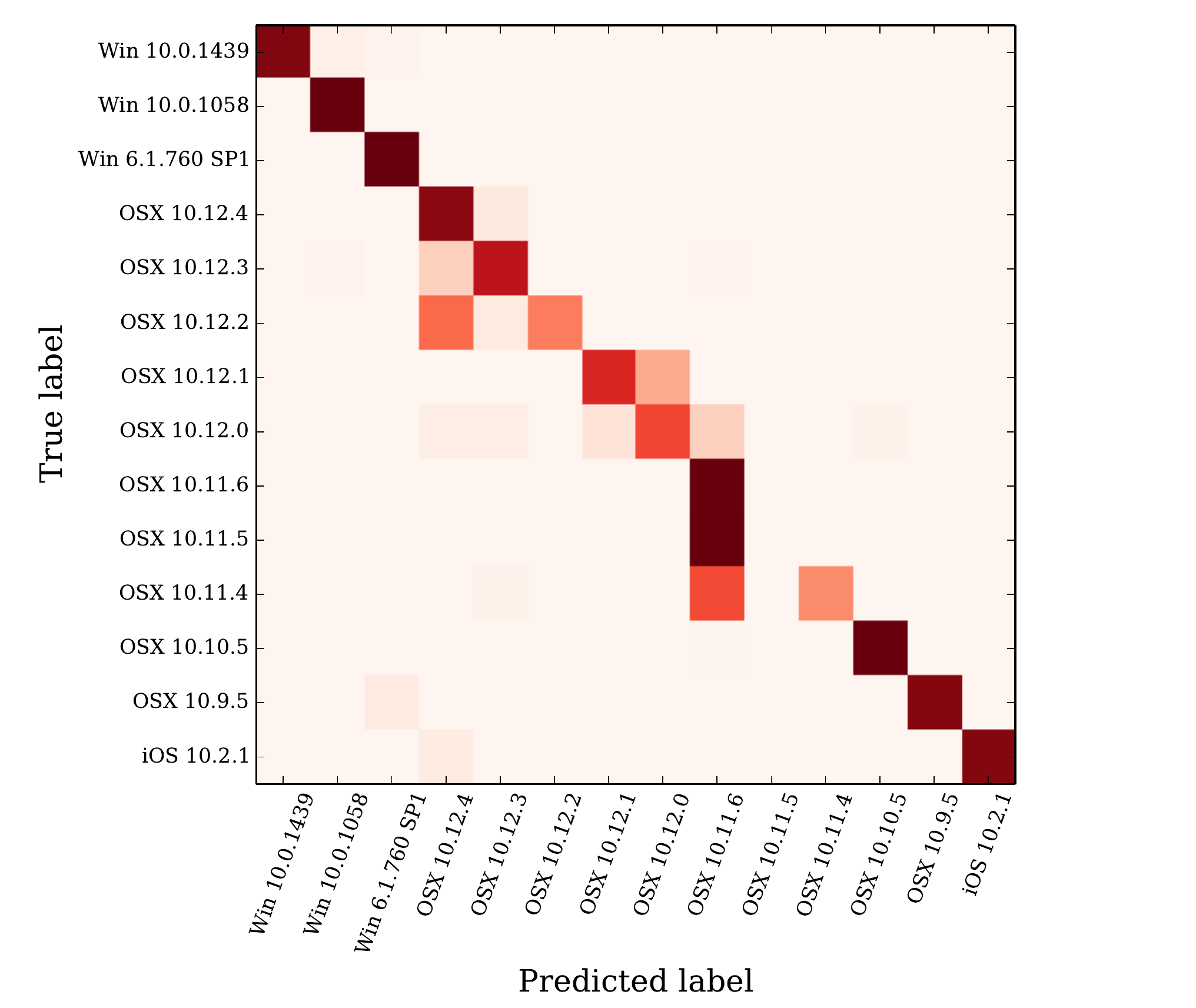}
   \caption{All, Accuracy=97.50\%}
   \label{fig:multi_flow_conf_all}
\end{subfigure}
\caption{Confusion matrices for the multi-session, random forest classifier.}
\label{fig:multi_flow_conf}
\end{figure*}

In contrast to treating each network flow independently, this experiment collects all fingerprints from an endpoint within a 60 minute window, and then classifies that window. We used a random forest model \cite{sklearn} to classify the windows, and used grid search and cross-validation on the training data to adjust the number of features per split and the depth of the trees. We also adjusted the number of trees in the forest, with 75 trees being optimal for nearly all problems. To ensure that the random forest had a minimal amount of training data for a specific OS type, we removed any OS type that had fewer than 10 active windows in the training dataset.

For each fingerprint type, we created a binary feature vector by defining a feature for each unique fingerprint that had at least 100 occurrences in the training dataset, and we set the binary feature to 1 if the associated fingerprint was observed in the window. We experimented with using a normalized count of each fingerprint, but this led to slightly inferior results. The TCP/IP feature vector had a length of 31, and, coincidentally, both the TLS and HTTP feature vectors had a length of 292. We also introduce a fourth model that leverages all available fingerprints by concatenating all three feature vectors; this feature vector had a length of 615. Table \ref{table:fingerprints_per_window} lists the average number of unique fingerprints observed in each 60 minute window. As one would expect, there are typically only 1-2 TCP/IP fingerprints and 3-4 HTTP fingerprints. TLS had the largest number of unique fingerprints in a window with an average over 10. This was in part due to small changes in the advertised extensions, but was mostly due to unique cipher suite offer vectors, which indicates that several unique applications are using TLS on the endpoint.

Figure \ref{fig:multi_flow_conf} presents the confusion matrices and total accuracy numbers for the four feature vector types. The multi-session approach significantly improves performance for TCP/IP and TLS, but provides similar performance to the single session model for HTTP. HTTP's performance was most likely due to less informative fingerprints, e.g., \texttt{Mac OS X 10.12}, dominating informative fingerprints, e.g., \texttt{Mac OS X 10.12.2}. %Feature pre-processing could help to alleviate this issue, but we leave this to future work.

In the case of TCP/IP, all of the improvements are due to better differentiation of Windows, which was more likely to use multiple TCP window sizes in the same 60 minute window. Using all fingerprints within a 60 minute window most improved the TLS model, $76.39\% \rightarrow 93.67\%$, which was due to the large number of unique fingerprints within a given window. Unsurprisingly, using all feature types gave the highest accuracy of the four models with $97.50\%$.

\begin{figure}
	\centering
   \includegraphics[scale=0.39]{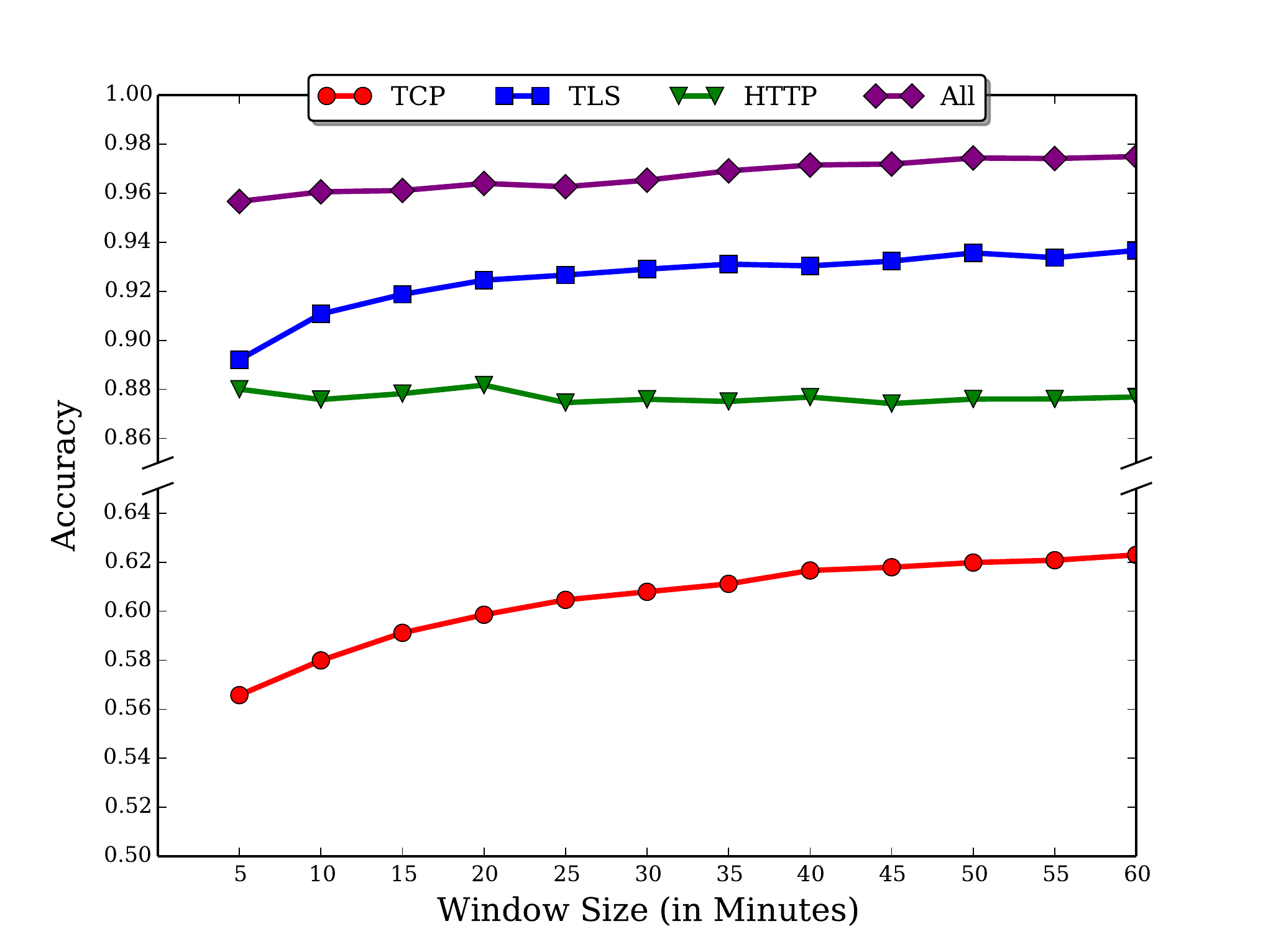}
	\caption{Accuracy given varying window sizes.}
	\label{fig:time_window}
\end{figure}

Figure \ref{fig:time_window} presents the effect of varying the window size from 5 to 60 minutes on the classifiers' performance. Both TLS and TCP significantly improve as the window size becomes larger, and start to plateau between 25 to 40 minutes. ``All" also steadily improves as the window size is increased, $95.67\% \rightarrow 97.50\%$. For reasons described above, HTTP's performance seems to be independent of the window size.

\subsection{General Categories}
\label{section:results:general_cats}

\bgroup
\def\arraystretch{1.2}
\begin{table}
\center
\begin{tabular}{l|c|c}
\hline
Classifier & Original & General \\
 & Categories & Categories \\
\hline
Single, TCP/IP & 55.54\% & 55.99\% \\
Single, TLS    & 76.39\% & 79.19\% \\
Single, HTTP   & 87.41\% & 93.42\% \\
Multi, TCP/IP  & 62.30\% & 63.12\% \\
Multi, TLS     & 93.67\% & 97.82\% \\
Multi, HTTP    & 87.69\% & 94.99\% \\
Multi, All     & 97.50\% & 99.40\% \\
\hline
\end{tabular}
\caption{General OS classifier accuracy.}
\label{table:general_categories_accuracy}
\end{table}
\egroup

While knowing the exact minor version number of an OS is beneficial, knowing just the major version number can be sufficient in some scenarios, and a classifier that identifies only the major version can obtain more accurate results. To examine these trade-offs, we performed the same experiments from the previous sections, but used a more general set of categories. After removing infrequent classes, the full label set was mapped to: $\lbrace\texttt{Win 10},$ $\texttt{Win 7/8},$ $\texttt{OS X 10.12},$ $\texttt{OS X 10.11},$ $\texttt{OS X 10.10},$ $\texttt{OS X 10.9},$ $\texttt{iOS 10}\rbrace$.

Table \ref{table:general_categories_accuracy} summarizes the performance gains from using the more general operating system categories. Most models and data types show a significant improvement. TCP/IP has only modest improvement, which is a result of misclassifying most major versions as either $\texttt{Win 7/8}$ or $\texttt{OS X 10.11}$, similar to the previous TCP/IP results. The windowed HTTP model begins to show improvements over the single session model, which is most likely due to the less informative fingerprints, e.g., \texttt{Mac OS X 10.12}, not impacting the predicted label. The windowed TLS model and the model that used all feature types showed improvements over the original categories, and had the most competitive accuracy numbers with $97.82\%$ and $99.40\%$, respectively.

\subsection{Detecting Vulnerable Operating Systems}
\label{section:results:vuln_cats}

Common vulnerabilities are collected and organized into central repositories \cite{cve_database}. After a vulnerability is disclosed, it is often quickly used to create exploits \cite{DBLP:conf/ccs/BilgeD12}. When network administrators need to identify vulnerable endpoints, a detector that can differentiate between vulnerable / not vulnerable is adequate. Similar to Section \ref{section:results:general_cats}, we perform this experiment by modifying the label set. We created two ``good" categories: $\lbrace\hspace{.2mm}\texttt{Win},$ $\texttt{OS X}\hspace{.2mm}\rbrace$ where $\texttt{Win} = \lbrace\hspace{.2mm}\texttt{Win 10.0.1439} \hspace{.2mm}\rbrace$ and $\texttt{OS X} = \lbrace\hspace{.2mm}\texttt{Mac OS X 10.12.5},\hspace{1mm}\texttt{Mac OS X 10.12.4\hspace{.2mm}}\rbrace$. There were also two vulnerable categories for \texttt{OS X} and \texttt{Win} that contained all other versions of those operating systems.

\begin{figure}
	\centering
   \includegraphics[scale=0.29]{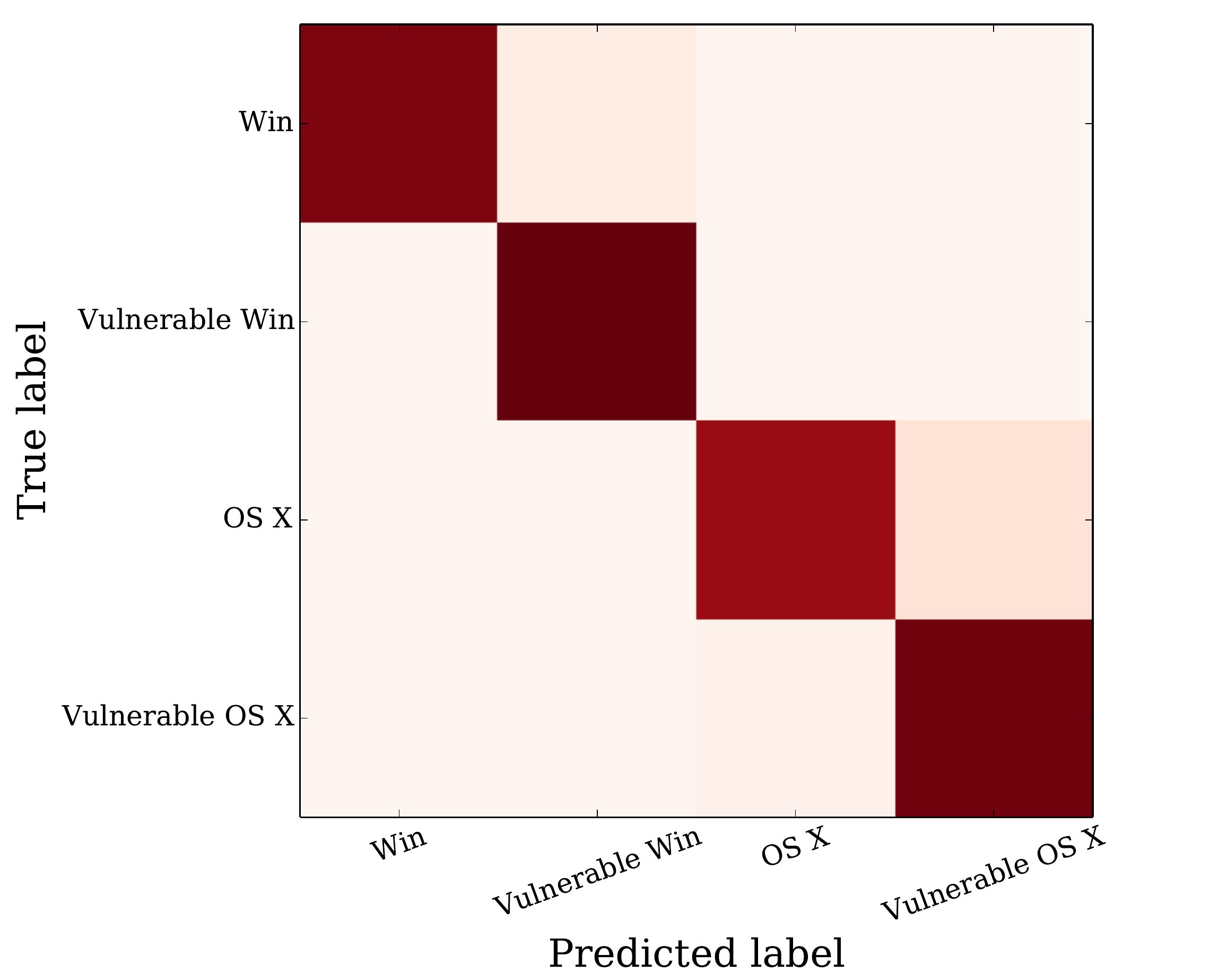}
	\caption{Confusion matrix for the multi-session, All classifier for vulnerable OS detection. Total accuracy was 98.05\%}
	\label{fig:multi_flow_conf_all_vuln}
\end{figure}

Figure \ref{fig:multi_flow_conf_all_vuln} presents the confusion matrix using all data types for this problem. The total accuracy of this model was 98.05\%. In the testing dataset, there were 7,083 \texttt{Vuln Win}, 2,407 \texttt{Vuln OS X}, 587 \texttt{Win}, and 1,045 \texttt{OS X} 60 minutes windows. There were 75 false negatives and 142 false positives. If false positives are an issue, cost-sensitive learning could be employed \cite{elkan2001foundations}.

%Figure \ref{fig:multi_flow_conf_all_vuln} presents the confusion matrix using all data types for this problem. The total accuracy of this model was 98.05\%. There were 7,083 \texttt{Vuln Win} and 2,407 \texttt{Vuln OS X} 60 minutes windows, and there 19 and 56 misclassifications for these categories, respectively. There were only 587 \texttt{Win} and 1,045 \texttt{OS X} 60 minutes windows, and the class imbalance for the \texttt{Win} categories contributed to the low number of false negatives. If a balanced dataset is not possible to obtain and false positives are an issue, cost-sensitive learning approaches could be employed \cite{elkan2001foundations}.

\subsection{Evasion}
\label{section:results:adversarial_learning}

Although the intention of this paper is to present methods that help network administrators defend and maintain a private network, these results could be used by attackers to fingerprint potential victims. In response to this threat, Albanese et.~al. presented tools to evade fingerprinting techniques that rewrite fields in network headers \cite{conf/cns/AlbaneseBJ15}. We now assess the performance of our system in the presence of such obfuscation techniques.

To test the feasibility of evasion, we developed and tested a straightforward obfuscation strategy. We created an obfuscation mapping that mapped all Windows hosts to the most prevalent Mac host, \texttt{Mac OS X 10.11.6}, and mapped all Mac/iOS hosts to the most prevalent version of Windows, \texttt{Win 6.1.760 SP1}. We maintained the original operating system categories, but would modify the fingerprints that we observed from a host to match those of the mapped host. The target fingerprints were chosen probabilistically to match the appropriate operating systems fingerprint distribution. 

We make the assumption that the device owner can control different percentages of the outgoing data, e.g., a user can typically choose their browser, but may not be able to influence the HTTP client used by an anti-virus agent. To mimic this assumption, we chose different percentages of a host's fingerprints to modify in a given 60 minute window, ranging from 0\% to 100\% with a step size of 25\%.  While more advanced obfuscation strategies may be possible, our tests highlight some of the key aspects of obfuscation against fingerprinting that uses multiple data types in a multi-session model.  %, and leave this research for future work.
%In all of these experiments,
The training data and categories were unmodified. 

\begin{figure}
	\centering
   \includegraphics[scale=0.39]{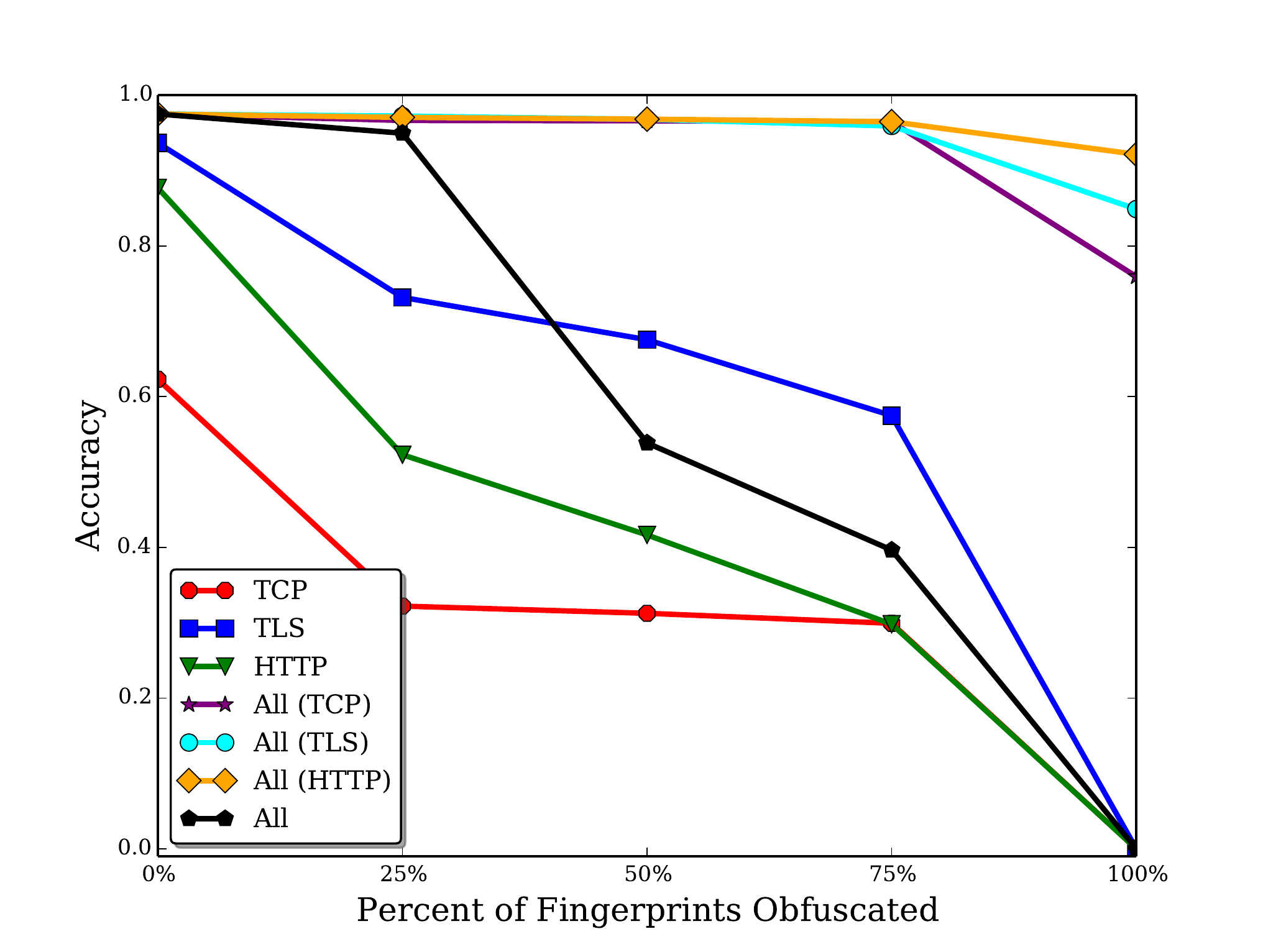}
	\caption{Accuracy of the classifiers based on sets of data features and different levels of obfuscation.}
	\label{fig:multi_flow_adv_decay}
\end{figure}

Figure \ref{fig:multi_flow_adv_decay} illustrates the performance of the multi-session classifiers when our obfuscation strategy is employed by all endpoints in the test dataset. The classifiers for all individual fingerprinting types show a substantial decrease in performance by modifying as little as 25\% of the test sample's fingerprints. TLS has the least amount of degradation, $93.67\% \rightarrow 73.17\%$, which is again due to the large number of TLS fingerprints in each 60 minute window. When all fingerprint types are modified equally, ``All" in Figure \ref{fig:multi_flow_adv_decay}, the classifier using all available data is robust against an obfuscation level of 25\%, $97.50\% \rightarrow 94.95\%$, but performs poorly at an obfuscation level of 50\% or higher.

We also tested the classifiers under the assumption that a device can only modify one fingerprint type, e.g., the results assuming that a device can only modify HTTP \texttt{User-Agent} strings is labeled as ``All (HTTP)" in Figure \ref{fig:multi_flow_adv_decay}. Under this assumption, the classifier using all three data types performed almost identically to the baseline model up to an obfuscation level of 75\%. At an obfuscation level of 100\%, the models did begin to show some weaknesses. At this level, TCP modifications had the most significant impact, decreasing the accuracy to 75.93\%. HTTP modifications had the least significant impact, only decreasing the accuracy to 92.15\%. The key point is that only modifying one fingerprint type or only modifying the fingerprints from certain applications will not successfully obfuscate a fingerprinting system that uses multiple, distinct fingerprint types over a long period of time.

\section{Discussion and Future Work}

The results put forth in this paper, specifically Section \ref{section:results:adversarial_learning}, have several important implications related to the Security/Privacy trade-off. As our results show, if a user can control 50\% or more of all fingerprint data types coming their device, evasion is possible. Unmanaged endpoints can reach this threshold by using operating systems that allow TCP stack modifications, only using carefully modified TLS libraries, only using HTTP clients that allows users to have full control over the \texttt{User-Agent} string and other headers, and keeping the number of applications that may violate these rules to a minimum. This level of control is nontrivial, but is possible if a user is concerned about privacy or a sensitive device needs to evade an attacker's reconnaissance.

That level of flexibility is not always possible in an enterprise network environment. In this setting, passive operating system detection can be used to identify vulnerable operating systems that need to be updated or quarantined. Endpoints on a corporate network often require particular operating systems and a large set of pre-installed applications, e.g., an anti-virus agent that a user cannot modify. In this situation, it is not generally possible for devices to control at least 50\% of the fingerprints. The device could create a large number of unique fingerprints, but this would be easily flagged as abnormal behavior. Our methods would perform well for enterprise network administrators that need additional visibility.

Unfortunately, our environment did not contain any Linux endpoints, which was a side effect of the VPN client, which we acquired ground truth from, not supporting Linux. We conjecture that the data features that we collect would make it possible to also distinguish Linux hosts with similar or higher accuracies. For instance, there are a set of TLS libraries with higher prevalence in Linux, e.g., \texttt{GnuTLS}, and there are also descriptive \texttt{User-Agent} strings. Given a properly labeled dataset with a sufficient number of target operating system examples, our approach is easily extensible.

Additional future work could include modifications to the data collection strategy that would allow one to gather labeled data from internal endpoints, which would remove the TCP MSS artifacts. In terms of feature pre-processing, we experimented with two algorithms to create the feature vectors from the individual fingerprints for the machine learning algorithms. The first, which had superior results and was used in our presented experiments, used any fingerprint that occurred at least 100 times in the training dataset, and the second used the top-$N\%$ of the fingerprints according to the fingerprints' normalized information gain, where $N \in \lbrace 10, 20, \ldots, 100 \rbrace$. Neither of these algorithms include any semantics about the specific feature types, e.g., \texttt{Mac OS X 10.12} could have high information gain, but lead to many false positives with the \texttt{10.12} minor version numbers. It would be interesting to explore more informative feature pre-processing metrics that incorporate this type of domain knowledge. Finally, assessing a number of different obfuscation strategies would help to tighten the bounds on the classifiers performance in this setting, e.g., alter the operating system obfuscation mapping to point to obscure or multiple operating systems.

%TCP/IP unmodified
%more feature pre-processing to eliminate low information gain fingerprints
%more advanced obfuscation strategies

\section{Ethical Considerations}
\label{section:ethical}

While collecting and analyzing the data needed for our experiments, we had to deal with highly confidential and sensitive data. We took all possible precautions to maintain the privacy of the end user, followed all institutional rules and procedures, and obtained the appropriate authorizations.  While collecting the data, all IP addresses and enterprise user names in HTTP header fields were anonymized via deterministic encryption. To obtain ground truth about the operating system types, a list of IP addresses were compiled from the full dataset, sent to a secure machine, and then deanonymized to facilitate querying the VPN logs for OS ground truth. The OS types were then associated with the anonymized IP address and a timestamp. The anonymized IP address/timestamp pair was then sent back to the machine that performed the experiments. The anonymized/deanonymized pairs were deleted immediately after ground truth was determined. 

\section{Conclusions}
Passive operating system fingerprinting generates useful intelligence for network defenders. By using TCP/IP, HTTP, and TLS features together in the multi-session model, accurate fingerprinting is possible, even to the level of minor version detection. A machine learning classifier can deal with the multitude of data features effectively, and we have shown that this approach provides better accuracy than single session fingerprints. The inclusion of TLS fingerprints for operating system identification is particularly important because HTTP is being replaced with the TLS-encrypted HTTPS protocol, and the traditional \texttt{User-Agent} strings will no longer be visible. The multi-session model presented in this paper enables one to easily add additional, distinct fingerprinting data types, which we have shown to be an important characteristic of an effective fingerprinting scheme.

Crucially, we demonstrated that a multi-session model based on TLS, HTTP, and TCP/IP can identify vulnerable operating systems with high accuracy, and that fingerprinting can be robust even when faced with levels of data feature obfuscation that could be seen on an enterprise network. In the more general setting, the multi-session model's combination of disparate data types, along with the fact that some data types are difficult to modify, allows the model to maintain performance even when 75\% of the features derived from individual data types are obfuscated. 

\section{Acknowledgements}

We would like to thank the Cisco Technology Fund for funding this work, and Greg Akers for his support. We would also like to thank Brian Long and Cisco CSIRT for their help in obtaining ground truth, and Philip Perricone and Bill Hudson for their contributions to Joy.

\bibliographystyle{plain}
\bibliography{cns}
%\bibliography{ieeesecpriv,rfc,mcgrew,johnq,pevnak,crypto}
%\bibliographystyle{IEEEtran}
%\bibliography{IEEEabrv,ieeesecpriv,rfc,mcgrew,johnq,pevnak,crypto}

%\appendices

\end{document}